# Escaping the Impossibility of Fairness: From Formal to Substantive Algorithmic Fairness

Ben Green[1]

**Abstract**
Efforts to promote equitable public policy with algorithms appear to be fundamentally constrained by the "impossibility of fairness" (an incompatibility between mathematical definitions of fairness). This technical limitation raises a central question about algorithmic fairness: How can computer scientists and policymakers support equitable policy reforms with algorithms? In this article, I argue that promoting justice with algorithms requires reforming the methodology of algorithmic fairness. First, I diagnose the problems of the current methodology for algorithmic fairness, which I call "formal algorithmic fairness." Because formal algorithmic fairness restricts analysis to isolated decision-making procedures, it leads to the impossibility of fairness and to models that exacerbate oppression despite appearing "fair." Second, I draw on theories of substantive equality from law and philosophy to propose an alternative methodology, which I call "substantive algorithmic fairness." Because substantive algorithmic fairness takes a more expansive scope of analysis, it enables an escape from the impossibility of fairness and provides a rigorous guide for alleviating injustice with algorithms. In sum, substantive algorithmic fairness presents a new direction for algorithmic fairness: away from formal mathematical models of "fair" decision-making and toward substantive evaluations of whether and how algorithms can promote justice in practice.

✉ Ben Green
bzgreen@umich.edu

1 University of Michigan, Ann Arbor, MI, USA

# 1 Introduction

## 1.1 Algorithmic Fairness and Its Discontents

Machine learning algorithms have become central components in many efforts to promote equitable public policy. In the face of widespread concerns about discriminatory institutions and decision-making processes, many policymakers, policy advocates, and scholars praise algorithms as critical tools for enhancing equality (115th United States Congress, 2017; Arnold Ventures, 2019; Eubanks, 2018; Porrino, 2017). To these proponents, algorithms overcome the cognitive limits and social biases of human decision-makers, enabling more objective and fair decisions (115th United States Congress, 2017; Arnold Ventures, 2019; Kleinberg et al., 2019; Miller, 2018; Sunstein, 2019). Thus, for instance, in light of concerns about the biases of judges, many court systems in the USA have adopted pretrial risk assessment algorithms as a central component of criminal justice reforms (Green, 2020; Koepke & Robinson, 2018; Porrino, 2017).

Undergirding these reform efforts is the burgeoning field of algorithmic fairness. Grounded primarily in computer science, algorithmic fairness applies the tools of algorithm design and analysis—in particular, an emphasis on formal mathematical reasoning (Green & Viljoen, 2020)—to fairness. The central components of algorithmic fairness are developing mathematical definitions of fair decision-making (Barocas et al., 2019), optimizing algorithms for these definitions (Feldman et al., 2015; Hardt et al., 2016), and auditing algorithms for violations of these definitions (Angwin et al., 2016; Obermeyer et al., 2019; Raji & Buolamwini, 2019).

In the context of policy reform efforts, algorithmic fairness methods are often employed to determine whether an algorithm is "fair" and, therefore, appropriate to use for making decisions. For instance, in settings such as pretrial adjudication and child welfare, debates about whether to employ algorithms hinge on evaluations of algorithmic fairness (Angwin et al., 2016; Chouldechova et al., 2018; Dieterich et al., 2016; Eubanks, 2018). Similarly, regulations of government algorithms often call for evaluations that test algorithms for biases (Brown, 2020; European Commission, 2021; Government of Canada, 2021; Le, 2021).

Yet as algorithmic fairness has risen in prominence, critical scholars have challenged its methods. Efforts to formulate mathematical definitions of fairness overlook the contextual and philosophical meanings of fairness (Binns, 2018; Green & Hu, 2018; Jacobs & Wallach, 2021; Lee et al., 2021; Selbst et al., 2019). Algorithmic fairness focuses on bad actors, individual axes of disadvantage, and a limited set of goods, thus "mirroring some of antidiscrimination discourse's most problematic tendencies" as a mechanism for achieving equality (Hoffmann, 2019). As a result, there is often a significant gap between mathematical evaluations of fairness and an algorithm's real-world impacts (Green & Viljoen, 2020). Algorithms that satisfy fairness standards often exacerbate oppression and legitimize unjust institutions (Davis et al., 2021; Green, 2020; Kalluri, 2020; Ochigame, 2020; Ochigame et al., 2018; Powles & Nissenbaum, 2018). In turn, some scholars have called for rejecting the frame of "fairness" altogether, proposing alternative frames of "justice" (Bui & Noble, 2020; Costanza-Chock, 2020; Green, 2018), "equity" (D'Ignazio & Klein, 2020), and "reparation" (Davis et al., 2021).

However, efforts to achieve algorithmic justice in practice are constrained by a fundamental technical limitation: the "impossibility of fairness." This result reveals that it is impossible for an algorithm to satisfy all desirable mathematical definitions of fair decision-making (Chouldechova, 2017; Kleinberg et al., 2016). An algorithm that is fair along one standard will inevitably be unfair along another standard.[1] Although no mathematical definition of algorithmic fairness fully encapsulates the philosophical notion of fairness or justice (Binns, 2018; Green & Hu, 2018; Jacobs & Wallach, 2021; Lee et al., 2021; Selbst et al., 2019), each definition captures a normatively desirable principle.

The impossibility of fairness appears to drastically limit the potential of algorithms to promote equitable public policy: any effort to improve decision-making using algorithms will violate at least one normatively desirable fairness principle. This result suggests that the best way for algorithm developers to promote fairness or justice in practice is to select some (limited) fairness definitions at the expense of others (Berk et al., 2018; Costanza-Chock, 2020; Davis et al., 2021; Kleinberg et al., 2019; Wong, 2020). Yet this strategy is fundamentally limited. As one article about algorithmic fairness concludes, "the tradeoff between […] different kinds of fairness has real bite" and means that "total fairness cannot be achieved" (Berk et al., 2018). Similarly, a proponent of algorithmic justice acknowledges that, because of the legal implications of these trade-offs, "it is highly unlikely that an algorithmic justice approach will advance" (Costanza-Chock, 2020).

The impossibility of fairness thus raises a central question about algorithmic fairness: How can computer scientists and policymakers support equitable policy reforms with algorithms? In this article, I argue that developing a rigorous strategy for algorithmic justice requires reforming the methodology of algorithmic fairness.

### 1.2 Article Overview: Methodological Reform

A methodology is "a body of methods, rules, and postulates employed by a discipline" (Merriam-Webster, 2021). A methodology provides a systematic language for comprehending and reasoning about the world, shaping how practitioners formulate problems and develop solutions to those problems. Problem formulation has both practical and normative stakes (Passi & Barocas, 2019). As philosopher John Dewey writes, "The way in which [a] problem is conceived decides what specific suggestions are entertained and which are dismissed" (Dewey, 1938). An inadequately conceived problem "cause[s] subsequent inquiry to be irrelevant or to go astray;" the remedy is to reformulate the problem (Dewey, 1938). Furthermore, as philosopher

---

[1] I will provide more detail on these fairness definitions and the impossibility of fairness in Section 2.

Elizabeth Anderson describes, "Sound political theories must be capable of representing normatively relevant political facts. If they can't represent certain injustices, then they can't help us identify them. If they can't represent the causes of certain injustices, then they can't help us identify solutions" (Anderson, 2009). In sum, if a methodology fails to account for normatively relevant facts and principles, it will generate problem formulations that yield unhelpful or unjust proposals for reform.

In the spirit of Dewey and Anderson, this article proposes methodological reforms so that algorithmic fairness can provide a more rigorous guide for promoting justice with algorithms. This goal involves two central tasks.

The first task is to diagnose why the current methodology for algorithmic fairness is flawed. I argue that the flaws of algorithmic fairness result from a significant methodological limitation: algorithmic fairness relies on a narrow frame of analysis restricted to specific decision points, in isolation from the context of those decisions.[2] I call this method "formal algorithmic fairness," as it aligns with formal equality (which emphasizes equal treatment for individuals based on their attributes or behavior at a particular decision point). Formal algorithmic fairness represents a systematic approach to problem formulation in which fairness is operationalized in terms of isolated decision-making processes. Because formal algorithmic fairness is conceived so narrowly, it yields a misguided and techno-centric reform strategy: enhance fairness by optimizing decision-making procedures with algorithms. These algorithmic interventions often exacerbate oppression and are constrained by the impossibility of fairness. Thus, formal algorithmic fairness leaves reformers in a bind: it appears that the only options are to adopt superficially "fair" algorithms or to reject algorithmic reforms, leaving the status quo in place.

The second task is to propose an alternative approach to algorithmic fairness that operationalizes a justice-oriented agenda for developing and applying algorithms. I call this new methodology "substantive algorithmic fairness," as it draws on legal and philosophical theories of substantive equality (which aim to eliminate social conditions of domination and oppression). My goal is not to incorporate substantive equality into a formal mathematical model: this strategy would fail to provide the necessary methodological shift (Green & Viljoen, 2020). Instead of treating fairness as a technical attribute of algorithms, substantive algorithmic fairness focuses on whether and how algorithms can promote equity in practice. Substantive algorithmic fairness thus suggests a two-pronged reform strategy that goes beyond striving to achieve formal equality within decision-making processes. First, reduce the upstream social disparities that feed into decision-making processes. Second, reduce the downstream harms that result for those judged unfavorably within decision-making processes. These strategies yield algorithmic interventions that escape from the impossibility of fairness and, in turn, can promote justice in practice.

---

[2] By decision points, I refer to the specific moments in which decisions are made about individuals. Examples include decisions about whether to release or detain pretrial defendants and decisions about whether to admit or reject college applicants.

## 2 The Impossibility of Fairness

In May 2016, journalists at *ProPublica* reported that a risk assessment algorithm used to judge pretrial defendants in Broward County, Florida was "biased against blacks" (Angwin et al., 2016). This algorithm, known as COMPAS, was created by the company Northpointe and is used by many court systems across the USA.[3] Like other pretrial risk assessments, COMPAS predicts the likelihood that pretrial defendants will recidivate; these predictions are presented to judges to inform their decisions to release or detain each defendant until their trial (Green, 2020; Koepke & Robinson, 2018). *ProPublica* found that, among defendants who were not arrested in the two years after being evaluated, Black defendants were 1.9 times more likely than white defendants to be misclassified by COMPAS as "high risk" (i.e., subjected to false positive predictions) (Angwin et al., 2016).

This report sparked significant debate about the use of COMPAS in pretrial adjudication. Tech critics responded to *ProPublica*'s article with outrage about racist algorithms (Doctorow, 2016; O'Neil, 2016). However, Northpointe and numerous academics defended COMPAS, arguing that *ProPublica* had focused on the wrong measure of algorithmic fairness (Corbett-Davies et al., 2017; Dieterich et al., 2016; Flores et al., 2016; Gong, 2016). These groups asserted that the proper standard of fairness is not whether false positive (and false negative) rates are the same for each race. Instead, they argued that the proper standard of fairness is whether risk scores imply the same probability of recidivism for each race. COMPAS satisfied this standard, suggesting that the tool was, in fact, fair.

This debate about COMPAS concerns two distinct definitions of algorithmic fairness. The first is "separation," which is satisfied if all groups subject to an algorithm's predictions experience the same false positive rate and the same false negative rate.[4] Separation expresses the idea that people who exhibit the same outcome should be treated similarly. *ProPublica* argued that COMPAS is biased because it violates separation: Black non-recidivists are more likely to be labeled "high risk" than white non-recidivists (Angwin et al., 2016).

The second notion of algorithmic fairness is "sufficiency," which is satisfied if, among those who receive a particular prediction, all groups exhibit the outcome being predicted at the same rate.[5] Sufficiency expresses the idea that people who are equally likely to exhibit the behavior of interest should be treated similarly. Northpointe and others argued that COMPAS is fair because it satisfies sufficiency: the label of "high risk" signifies a similar probability of recidivism for both Black and white defendants (Corbett-Davies et al., 2017; Dieterich et al., 2016; Flores et al., 2016; Gong, 2016). Sufficiency is the most widely used notion of algorithmic fairness, particularly because machine learning models typically satisfy this principle by default (Barocas et al., 2019).

---

[3] COMPAS stands for Correctional Offender Management Profiling for Alternative Sanctions. Northpointe has since been renamed Equivant.

[4] Separation is aligned with fairness criteria such as error rate balance and balance for the positive/negative class.

[5] Sufficiency is aligned with fairness criteria such as calibration and predictive parity.

The COMPAS debate raised a fundamental question for algorithmic fairness: can an algorithm simultaneously satisfy both separation and sufficiency? As computer scientists soon discovered, the answer is no: there is an inevitable tension between these definitions of fairness (Angwin & Larson, 2016; Barocas et al., 2019; Chouldechova, 2017; Kleinberg et al., 2016). This result is known as the "impossibility of fairness." The only exceptions to the impossibility of fairness involve two exceedingly rare scenarios: the algorithm makes predictions with perfect accuracy, or all groups exhibit the outcome being predicted at the same "base rate" (Kleinberg et al., 2016).

The impossibility of fairness reflects a harsh and intractable dilemma facing efforts to promote equality using algorithms (Berk et al., 2018). This dilemma is especially troubling in public policy, where algorithms are typically adopted to enhance the fairness of discrete decision-making processes. In these settings, the statistical fairness measures in tension are salient and often grounded by law. The impossibility of fairness raises a particular challenge for proponents of algorithmic justice: because their proposals involve violating sufficiency in favor of alternate measures (Costanza-Chock, 2020; Davis et al., 2021), such attempts would generally be barred by antidiscrimination law (Corbett-Davies et al., 2017; Costanza-Chock, 2020; Hellman, 2020).

Efforts to promote algorithmic fairness and algorithmic justice operate within the constraints imposed by the impossibility of fairness. The impossibility of fairness suggests that reformers can either (a) choose a single fairness definition at the expense of others or (b) rigorously balance the tradeoffs between multiple definitions (Berk et al., 2018; Costanza-Chock, 2020; Davis et al., 2021; Kleinberg et al., 2019; Wong, 2020). Yet as I will describe in Section 4, both of these responses lead to narrow reforms that uphold unjust social conditions and institutions. Developing a positive agenda for algorithmic justice requires finding a way to develop and apply algorithms without confronting the impossibility of fairness.

## 3 Lessons from Egalitarian Theory

To inform the evolution toward an algorithmic fairness methodology that promotes justice and escapes from the impossibility of fairness, I turn to egalitarian theory. Broadly speaking, "Egalitarian doctrines tend to rest on a background idea that all human persons are equal in fundamental worth or moral status" (Arneson, 2013). Although fairness and equality are complex and contested concepts, both share a central concern with comparing the treatment or conditions across individuals or groups, emphasizing the normative value of some form of parity (Arneson, 2013; Gosepath, 2021; Minow, 2021). Indeed, many definitions of algorithmic fairness explicitly reference equality (Barocas et al., 2019; Berk et al., 2018). Furthermore, egalitarian scholars have confronted many questions that overlap with central debates in algorithmic fairness (Binns, 2018; Lee et al., 2021).

### 3.1 Formal and Substantive Equality

Just as algorithmic fairness confronts narrow formulations of fairness, egalitarian theorists have confronted narrow formulations of equality. In response, some egalitarian thinkers have devised more expansive formulations of equality that provide a better guide for ameliorating oppression.

A central tension in egalitarian theory is between "formal" and "substantive" equality. Formal equality asserts, "When two persons have equal status in at least one normatively relevant respect, they must be treated equally with regard in this respect. This is the generally accepted *formal* equality principle that Aristotle articulated […]: 'treat like cases as like'" (Gosepath, 2021). In practice, formal equality typically refers to a "fair contest" in which everyone is judged according to the same standard, based only on their characteristics at the moment of decision-making (Fishkin, 2014). In the USA, disparate treatment law is grounded in formal equality, attempting to ensure that people are not treated differently based on protected attributes such as race and gender.

Despite being widely adopted, formal equality suffers from a methodological limitation. Because formal equality restricts analysis to specific decision points, it cannot account for the inequalities that often surround those decision points. Formal equality is therefore prone to reproducing existing patterns of injustice. For instance, a formal equality approach to college admissions would evaluate all applicants based solely on their academic qualifications (e.g., grades and test scores). As long as decisions are based on accurate evaluations and applicants with similar qualifications are treated similarly, formal equality would be satisfied. Yet because of racial inequalities in educational opportunities (EdBuild, 2019), evaluating all students according to a uniform standard would perpetuate racial inequality.

The limits of formal equality have led many scholars to develop an alternative: substantive equality. Substantive equality "repudiate[s] the Aristotelian 'likes alike, unlikes unalike' approach […] and replaces it with a substantive test of historical disadvantage" (MacKinnon, 2011). In particular, substantive equality is oriented toward identifying and remediating social hierarchies: "social relation[s] of rank ordering, typically on a group or categorical basis," that generate disparities in social and material resources (MacKinnon, 2011). Social hierarchies refer to caste-like arrangements in which dominant groups (e.g., white people, men) treat other groups (e.g., Black people, women) as inferior and subordinate. Substantive equality scholars thus advocate for abolishing social conditions that facilitate domination and oppression (Anderson, 1999; MacKinnon, 2011; Young, 1990). Substantive equality calls instead for institutional conditions that support self-determination and self-development, creating communities in which each person is treated as having equal moral worth (Anderson, 1999; MacKinnon, 2011; Young, 1990).[6] In the USA, disparate impact law is grounded in substantive equality (albeit partially (MacKinnon, 2011)), attempting to ensure that formally neutral rules do not disproportionately burden historically marginalized groups.

By recognizing that "oppression […] by definition is socially imposed" (Anderson, 1999), substantive equality expands the scope of analysis beyond isolated decision points to include social relationships and institutional arrangements. Through the lens of substantive equality, "material and dignitary deprivations and violations are substantive indications and consequences of [social] hierarchy, but it is the hierarchy itself that defines the core inequality problem" (MacKinnon, 2011). Thus, when confronted with instances of inequality, "A substantive equality approach […] begins by asking, what is the substance of this particular inequality, and are these facts an instance of that substance?" (MacKinnon, 2011). This emphasis on social hierarchies and institutional arrangements enables substantive equality to suggest reform strategies that challenge rather than reproduce existing patterns of injustice.

### 3.2 Substantive Approaches to Escaping Equality Dilemmas

Substantive equality is particularly helpful for dealing with dilemmas between competing approaches to equality. Just as algorithmic fairness confronts the impossibility of fairness, egalitarian theorists have confronted similar tensions between notions of equality. In response, some egalitarian thinkers have devised reform strategies that break free from these dilemmas.

In order to glean insights about how algorithmic fairness can escape the impossibility of fairness, I turn to three complementary substantive equality approaches for analyzing and escaping from equality dilemmas:

- In developing her theory of "democratic equality," philosopher Elizabeth Anderson responds to a "dilemma" that arises in luck egalitarianism (Anderson, 1999).[7] On the one hand, not providing aid to the disadvantaged means blaming individuals for their misfortune. On the other hand, providing special treatment to individuals on account of their inferiority means expressing contempt for the disadvantaged.

---

[6] It is worth making two clarifications about substantive equality. First, asserting that every person has equal moral worth does not require the assumption that every person has equal talents or virtue. Substantive equality focuses on how society responds to differences in people's attributes and capabilities (e.g., skin color, sex, physical ability). The aim is not to ensure that everyone has the same attributes and capabilities. Instead, substantive equality asserts that everyone must be treated with equal respect regardless of their attributes and capabilities (Anderson, 1999; MacKinnon, 2011; Minow, 1991). Second, it is important to distinguish between social hierarchies (i.e., caste-like social arrangements) and other forms of hierarchy in enabling productive collective activities. Substantive equality forbids the former but permits the latter (under certain conditions). In organizations such as companies and social movements, hierarchies of authority can enhance efficiency and productivity (Anderson, 2017; Tufekci, 2018). However, higher officeholders should not be able to wield arbitrary and unaccountable power over others. Instead, those in lower positions of authority should have standing to participate in decision-making and hold officeholders accountable (e.g., through the ability to elect and sanction officeholders) (Anderson, 2017).

[7] Luck egalitarianism advocates compensating people for inequalities that result from misfortunate but not inequalities that result from choice (Anderson, 1999; Arneson, 2013).

• In developing her "social-relations approach" to equality, legal scholar Martha Minow engages with the "dilemma of difference" that arises in legal efforts to deal with differences between individuals (Minow, 1991). On the one hand, giving similar treatment to everyone regardless of their circumstances can "freeze in place the past consequences of differences." On the other hand, giving special treatment to those deemed "different" risks entrenching and stigmatizing that difference.

• In developing his theory of "opportunity pluralism," legal scholar Joseph Fishkin addresses the "zero-sum struggles" that arise in efforts to promote equal opportunity (Fishkin, 2014). On the one hand, judging people for an opportunity based solely on their performance or attributes at a particular moment in time (i.e., a "fair contest") perpetuates inequalities. On the other hand, approaches that attempt to account for existing inequalities (such as Rawlsian equal opportunity and luck egalitarianism) fail to create a truly level playing field and prompt "extraordinarily contentious" debates.

The equality dilemmas analyzed by Anderson, Minow, and Fishkin resemble the impossibility of fairness. In all of these cases, efforts to promote equality are impaired by a seemingly inescapable, zero-sum tension between notions of equality. If we treat everyone following a uniform standard (akin to sufficiency), we risk reproducing inequality. But if we provide special treatment to the disadvantaged (akin to separation), we might stigmatize the disadvantaged and still fail to achieve greater equality. It thus appears difficult—if not impossible—to meaningfully advance equality. As Minow notes, "Dilemmas of difference appear unresolvable" (Minow, 1991). In turn, "decisionmakers may become paralyzed with inaction" (Minow, 1991). At best, decision-makers appear to be left with a zero-sum trade-off between competing notions of equality. Yet as Fishkin writes, "If […] zero-sum tradeoffs are the primary tools of equal opportunity policy, then trench warfare is a certainty, and any successes will be incremental" (Fishkin, 2014).

In the face of these challenges, Anderson, Minow, and Fishkin provide methodological accounts of how to escape from these dilemmas. Each scholar reveals that their dilemma is not intractable. Instead, each dilemma only appears intractable if one analyzes inequality through a narrow lens, which restricts the range of possible remedies. Expanding the frame of analysis sheds new light on the problems of inequality and yields two reform strategies that escape these equality dilemmas.

The first substantive approach to escaping equality dilemmas is what I call the "relational response": reduce social and material disparities grounded in social hierarchy. Anderson and Minow broaden the analysis of equality from unequal distributions of goods and traits to social relations (Anderson, 1999; Minow, 1991). From this perspective, the problem of inequality is not that some people are inherently "different" from others or merely that some people have more of a particular good than others. An additional problem is that dominant groups have arranged many institutions and norms in ways that distribute goods and capabilities unequally along lines of social hierarchy (Anderson, 1999; Minow, 1991). Definitions of which attributes make someone normal, desirable, and worthy of support often reflect the interests of those in power (Minow, 1991).

The relational response therefore operates upstream from a given decision-making process. Reform should not simply provide special treatment to "different" or "inferior" individuals, accepting existing social relations as neutral and static. More broadly, reform should mitigate the extent to which oppressed groups disproportionately exhibit the attributes deemed "negative" within a given decision-making process. If norms and policies did not translate differences between people into disparities in normatively significant attributes, then decision-makers would not be confronted with the dilemma between treating everyone the same and providing special treatment.

The second substantive approach to escaping equality dilemmas is what I call the "structural response": reduce the scope and stakes of decisions that act on social disparities. Fishkin broadens the analysis of equality from individual competitions to the entire structure of opportunities. From this perspective, the problem of inequality is not merely that groups face vastly different development opportunities, making it impossible to create fair contests between all individuals. An additional problem is that opportunities are structured around a small number of "zero-sum, high-stakes competitions" (Fishkin, 2014). These competitions typically hinge on attributes that are unequally distributed across groups due to oppression, thus compounding existing disadvantage and raising the stakes of equality dilemmas.[8]

The structural response therefore operates downstream from a given decision-making process. Reform should not simply help some disadvantaged individuals receive favorable decisions through special treatment, accepting the structure of opportunities as given. More broadly, reform should "renovate the structure" (Fishkin, 2014) of decisions to limit the extent to which a given decision-making process punishes individuals who exhibit the attributes deemed "negative." If decision-making structures did not harm individuals who are judged negatively, then the dilemma between treating everyone the same and providing special treatment would have dramatically lower stakes.

### 3.2.1 Case Study: College Admissions

The relational and structural responses provide complementary substantive interventions when dealing with decision-making processes that perpetuate or exacerbate inequities. Consider the example of racial inequality in US college admissions. Debates about admissions decisions frame decision-making as a zero-sum choice between two options: should colleges evaluate all students based on a uniform standard (i.e., formal equality) or should they provide underprivileged students with special treatment (i.e., affirmative action)? The equality dilemma presented by this framing leads to heated lawsuits and political organizing (Fishkin, 2014; Lehmann, 2021; Minow, 1991). As a result, affirmative action in admissions has become a particularly fraught and frustrating terrain for progressives hoping to combat racial inequality (Fishkin, 2014; Lehmann, 2021; Leonhardt, 2020). The relational and structural responses suggest ways out of this tension.

---

[8] For instance, oppressed groups are generally less qualified to succeed in competitions for beneficial opportunities such as jobs, making hiring decisions particularly consequential and contentious.

The relational response would aim to decouple the link between race and educational achievement. Through a relational lens, debates about admission to colleges and universities are contentious not only because of concerns about biased decision-makers, but also because Black and white students receive drastically different levels of educational opportunities. Education policies in the USA are significantly shaped by racial hierarchy, such that white students (on average) achieve better educational performance than Black students (EdBuild, 2019; Ewing, 2018; Rothstein, 2015; Sharkey, 2013; Smith & Reeves, 2020). The relational response suggests one escape from the equality dilemma in this setting: reduce the racial disparity in educational attainment. Achieving this goal requires altering education policies to distribute resources more equitably and altering definitions of academic performance to prioritize skills beyond those typically promoted by dominant groups. By reducing the racial gap in academic performance, such reforms would reduce the extent to which college admissions decisions confront a dilemma between formal equality and affirmative action.

The structural response would aim to decouple the link between educational achievement and future life chances. Through a structural lens, debates about admission to colleges and universities are contentious not only because of racial inequities in educational attainment, but also because admission provides a particularly reliable pathway to high social status and material comfort. The significance of college admissions decisions makes disparities in primary and secondary education particularly consequential for determining future life outcomes. The structural response suggests an escape from the equality dilemma in this setting: lower the stakes of college admissions decisions. Achieving this goal requires altering the structure of opportunities to create more paths for people to lead comfortable and fulfilling lives without a college degree. By making college admissions less determinative of future life outcomes, such reforms would reduce the downstream harms of disparities in educational opportunities, thus lowering the stakes of the dilemma between formal equality and affirmative action.

These ideas from egalitarian theory have important lessons for algorithmic fairness. As I will describe in the following section, the current approach to algorithmic fairness is grounded in formal equality and shares many of formal equality's limits. This analysis suggests the need for an alternative approach grounded in substantive equality, which I will present in Section 5. The relational and structural responses from substantive equality suggest new strategies for how algorithmic fairness can escape the impossibility of fairness and better alleviate social hierarchies.

## 4 Formal Algorithmic Fairness: Navigating the Impossibility of Fairness

This section focuses on the first task of reforming algorithmic fairness: diagnosing the limits of algorithmic fairness as a guide for promoting equitable public policy. I characterize the dominant method of algorithmic fairness as "formal algorithmic fairness." Akin to formal equality, formal algorithmic fairness limits analysis to the functioning of algorithms at particular decision points. When confronted with concerns about discriminatory decision-making, formal algorithmic fairness formulates the problem in terms of only the inputs and outputs of the decision point in question. Fairness is therefore defined as technical attribute of algorithms.

Due to this narrow frame of analysis, formal algorithmic fairness suffers from many of the same methodological limits as formal equality. In this section, using pretrial risk assessments in the USA as a case study, I consider the two main responses to the impossibility of fairness that arise within formal algorithmic fairness. Interrogating these responses through the lens of substantive equality reveals how reforms that appear fair within formal algorithmic fairness can actually reproduce injustice.

### 4.1 The Fair Contest Response: Reproducing Inequity

The first formal algorithmic fairness response to the impossibility of fairness is what I call the "fair contest response." Most critiques of *ProPublica*'s COMPAS report asserted that the proper definition of algorithmic fairness is sufficiency (which COMPAS satisfies) rather than separation (which COMPAS violates) (Corbett-Davies et al., 2017; Dieterich et al., 2016; Flores et al., 2016; Gong, 2016). This response applies the logic of a fair contest: defendants should be evaluated based solely on their likelihood to recidivate. On this logic, COMPAS is fair and produces a higher false positive rate for Black defendants simply because they are more likely to recidivate. As Northpointe explained, the violation of separation presented by *ProPublica* "does *not* show evidence of bias, but rather is a natural consequence of using unbiased scoring rules for groups that happen to have different distributions of scores" (Dieterich et al., 2016).

The fair contest response asserts that fairness entails making decisions about people based solely on their likelihood to exhibit a particular outcome of interest. Under this logic, algorithmic bias is a problem of systematic misrepresentation (e.g., over-predicting the risk of Black defendants relative to the ground truth). Therefore, the best way to advance algorithmic fairness is to increase prediction accuracy and ensure that decisions are based on accurate judgments about each individual (Hellman, 2020; Kleinberg et al., 2019).

Because of this narrow focus on the decision-making process, the fair contest response fails to account for—and thus reproduces—broader patterns of injustice. First, the fair contest response treats risk as an intrinsic and neutral attribute of individuals. This response naturalizes group differences in risk that are the product of oppression. In the case of risk assessments, Black and white defendants do not just "happen to have different distributions of scores," as adherents of sufficiency assert (Dieterich et al., 2016). Instead, past and present discrimination has created social conditions in the USA in which Black people are empirically at higher risk to commit crimes (Butler, 2017; Cooper & Smith, 2011; Sampson et al., 2005).[9]

---

[9] This disparity results from oppression rather than from differences in inherent criminality (Muhammad, 2011). Furthermore, the disparity is true above and beyond racial disparities in arrest and enforcement patterns (i.e., measurement bias).

Second, the fair contest response ignores the consequences of the policies that the algorithm facilitates. When a risk assessment labels a defendant "high risk," that person is likely to be detained in jail until their trial. This practice of detaining defendants due to their crime risk, known as "preventative detention," has been critiqued as violating human rights, undermining the rights of the accused, and exacerbating mass incarceration (Baradaran, 2011; Koepke & Robinson, 2018; United States Supreme Court, 1987). Pretrial detention imposes severe costs on defendants, including the loss of freedom, an increased likelihood of conviction, and a reduction in future employment (Dobbie et al., 2018).

By failing to account for the social hierarchies and harmful policies associated with pretrial decision-making, the fair contest response suggests a reform strategy in which even the best-case scenario—a perfectly accurate risk assessment—would perpetuate racial inequity.[10] Because Black defendants recidivate at higher rates than white defendants (Cooper & Smith, 2011; Flores et al., 2016; Larson et al., 2016; Sampson et al., 2005), a perfect risk assessment will accurately label a higher proportion of Black defendants as "high risk" (after all, if data is collected about an unequal society, then an accurate algorithm trained on that data will reflect those unequal conditions). To the extent that these predictions direct pretrial decisions, this risk assessment would lead to a higher pretrial detention rate for Black defendants than white defendants. This would, in effect, punish Black communities for having been unjustly subjected to criminogenic circumstances in the first place, while providing the appearance of fairness. Thus, although a perfect risk assessment may help some Black defendants who are low risk but could be stereotyped as high risk, it would also naturalize the fact that many Black defendants actually are high risk and become incarcerated as a result.

### 4.2 The Formalism Response: Constraining Reform

The second formal algorithmic fairness response to the impossibility of fairness is what I call the "formalism response." Instead of choosing a single fairness metric, the formalism response focuses on balancing the tradeoffs between the competing metrics. Proponents of this response argue that the formalism of algorithms reveals and clarifies the difficult tradeoffs between notions of fairness that might otherwise remain opaque and unarticulated (Barocas et al., 2019; Berk et al., 2018; Kleinberg et al., 2019; Ligett, 2021; Sunstein, 2019). Under this view, algorithms can "be a positive force for social justice" because they "let us precisely *quantify tradeoffs* among society's different goals" and "force us to make more explicit judgments about underlying principles" (Kleinberg et al., 2019).[11]

---

[10] Because this risk assessment makes perfect predictions, it would satisfy both sufficiency and separation (Kleinberg et al., 2016). Fairness metrics that are always satisfied by a perfect classifier have been labeled "bias preserving," as they take existing social conditions as a neutral baseline (Wachter et al., 2021).

[11] The formalism response is inclusive of the fair contest response: after considering the tradeoffs, one could determine that an algorithm should be optimized for sufficiency. The formalism response can also account for other tradeoffs, such as the tension between fairness and accuracy. As with the fair contest response, a perfect risk assessment represents the best-case scenario, as it eliminates these tradeoffs.

Although the formalism response provides mathematical rigor about tradeoffs within particular decision points, it suggests a constrained and technocentric reform strategy. First, the formalism response leaves us stuck making a zero-sum choice between two limited notions of fairness. Although separation may appear to be a desirable alternative to sufficiency, it has several shortcomings. Creating an explicitly higher risk threshold for Black defendants would violate disparate treatment law in many instances (Corbett-Davies et al., 2017; Hellman, 2020). Furthermore, although a lack of separation demonstrates that different groups face disparate burdens from mistaken judgments (Chouldechova, 2017; Hellman, 2020), separation does not prevent the injustices associated with accurate predictions (Wachter et al., 2021). As the perfect pretrial risk assessment described in Section 4.1 demonstrates, an algorithm can satisfy separation while still reproducing racial hierarchy.

Second, the formalism response obscures pathways for systematic reform. Research on fairness in risk assessments explicitly takes structural reforms off the table at the outset of analysis, placing racial disparities outside the scope of fairness and the responsibility of developers (Chouldechova, 2017; Corbett-Davies et al., 2017; Kleinberg et al., 2019). This lens presumes that the only possible actions are to release and detain defendants, suggesting that reforms are limited to balancing the tradeoffs between values when deciding whom to release and detain. Following this logic, the formalism response suggests that the only possible (or, at least, pertinent) alternative to the status quo is to optimize specific decision-making processes using algorithms (Berk et al., 2018; Kleinberg et al., 2019; Miller, 2018). However, this approach is fundamentally limited as a strategy for achieving equitable public policy: efforts to remediate inequality that reform decision-making procedures alone often obscure and entrench the actual sources of oppression (Kahn, 2017; Murakawa, 2014). Implementing pretrial risk assessments thus legitimizes preventative detention as the appropriate response to high-risk defendants and hinders efforts to promote less carceral alternatives (Green, 2020).

In fact, the narrow purview of the formalism response is what makes the tension between sufficiency and separation appear to be such an intractable and troubling dilemma. What is strictly "impossible" is simultaneously satisfying all mathematical definitions of fairness when making decisions about individuals in an unequal society. However, because it relies on a highly constrained theory of social change limited to isolated decision points, the formalism response magnifies the stakes of this mathematical incompatibility. When all other aspects of society are treated as static or irrelevant, mathematical definitions of fair decision-making come to represent "total fairness" (Berk et al., 2018). Within this limited frame of analysis, the mathematical constraint on fair decision-making becomes a fundamental "impossibility of fairness," suggesting inescapable constraints on equality-enhancing reforms. In other words, the formalism response provides clarity only within a narrow scope of analysis that obscures and impedes action toward substantive reforms.

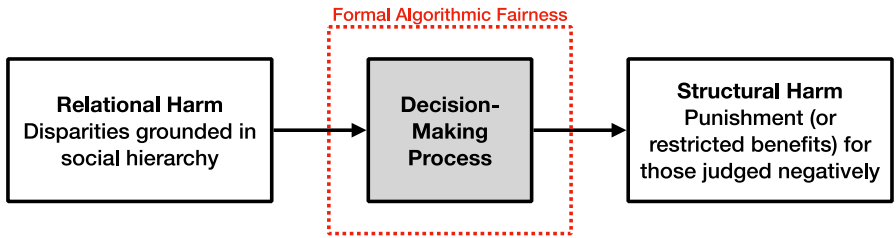

**Fig. 1** Flowchart depicting the causes of the impossibility of fairness and the limits of formal algorithmic fairness. Upstream of a decision-making process, the relational harm of group disparities creates the impossibility of fairness. If the relevant social groups exhibited the outcome of interest at the same rate, then it would be possible for the decision-making process to satisfy both separation and sufficiency with an imperfect algorithm. Downstream of a decision-making process, the structural harm of punishing individuals who exhibit the relevant "negative" attributes causes the impossibility of fairness to entrench injustice. If the decision-making process did not punish people who are judged negatively, then the group disparities would not lead to further harm. Because formal algorithmic fairness limits analysis to the decision-making process, it is unable to identify and account for relational and structural harms.

### 4.3 The Methodological Limits of Formal Algorithmic Fairness

This substantive equality analysis sheds light on the impossibility of fairness and the limits of formal algorithmic fairness. Debates and consternation about the impossibility of fairness arise when making decisions in which (a) an oppressed group disproportionately exhibits the attributes deemed "negative" in the given context (e.g., indicators of high crime risk), and (b) policy punishes (or restricts benefits to) individuals who exhibit these negative attributes (e.g., pretrial detention). When these relational and structural harms are present, any attempt to improve decision-making with an algorithm will confront the impossibility of fairness. Figure 1 depicts how relational and structural harms surround decision-making processes to make the impossibility of fairness such a troubling dilemma for algorithmic decision-making. If there were no social hierarchies or if consequential decisions did not exacerbate social hierarchies, then the impossibility of fairness would not arise (or, at least, would not be so concerning).

Figure 1 elucidates why formal algorithmic fairness is methodologically incapable of promoting justice in policy settings with relational and structural inequality. The fundamental problem lies with "[t]he way in which the problem is conceived" (Dewey, 1938): formal algorithmic fairness restricts analysis to isolated decision points. This limited scope means that formal algorithmic fairness "can't represent the causes of certain injustices" (Anderson, 2009). The fair contest and formalism responses both fail to account for relational and structural harms. In turn, formal algorithmic fairness "can't help us identify solutions" that address injustices (Anderson, 2009). Although the fair contest and formalism responses yield slightly different proposals, both responses suggest techno-centric reforms that entrench oppression and are trapped by the impossibility of fairness. Thus, in order to develop a positive agenda for algorithmic justice, it is necessary to develop a new methodology for algorithmic fairness that incorporates relational and structural considerations into the scope of analysis.

## 5 Substantive Algorithmic Fairness: Escaping the Impossibility of Fairness

Given the methodological limits of formal algorithmic fairness, this section focuses on the second task of reforming algorithmic fairness: proposing an alternative approach that operationalizes a social justice orientation into algorithmic fairness. As an alternative to formal algorithmic fairness, I propose a methodology of "substantive algorithmic fairness." Substantive algorithmic fairness is not a methodology for incorporating substantive equality into a formal mathematical model. That approach would narrow and distort the concept. Instead, substantive algorithmic fairness follows the approach of "algorithmic realism" (Green & Viljoen, 2020), expanding the scope of analysis to encompass the relational and structural considerations that surround particular decision points. Substantive algorithmic fairness does not entirely reject formal algorithmic fairness, however. Instead, it represents an expansion of algorithmic fairness methods, adopting substantive equality tools to reason about when formal algorithmic fairness is (and is not) appropriate.

Because of its broad frame of analysis, substantive algorithmic fairness accounts for the relational and structural harms at the heart of the impossibility of fairness. In doing so, substantive algorithmic fairness provides a guide for using algorithms to promote equitable public policy without being constrained by the impossibility of fairness. Substantive algorithmic fairness proposes that reforms should target relational and structural inequalities, not just the precise mechanisms of decision-making. This method therefore prompts advocates to interpret the impossibility of fairness not as a dead end or fundamental constraint on reform, but as a suggestion to consider a broader reform strategy. The proper response to the impossibility of fairness is not to tinker within the contours of this intractable dilemma, but to reform the relational and structural harms that produce the dilemma.

This section proceeds in three parts. First, I describe the general principles of substantive algorithmic fairness. Second, I apply substantive algorithmic fairness to pretrial reform. Third, I describe the next steps and challenges associated with implementing substantive algorithmic fairness in practice.

### 5.1 The Substantive Algorithmic Fairness Approach to Reform

As with formal algorithmic fairness, the starting point for reform in substantive algorithmic fairness is concern about discrimination or inequality within a particular decision-making process. Drawing on the substantive equality approaches introduced in Section 3, substantive algorithmic fairness presents a three-step strategy for promoting equality in such scenarios. The flowchart in Fig. 2 provides a guide for implementing these steps. This flowchart translates substantive equality goals into concrete questions for computer scientists, policymakers, and others to consider before developing and implementing an algorithm. In turn, the flowchart can direct reform efforts away from the narrow and techno-centric reforms typically suggested by formal algorithmic fairness. Nonetheless, the flowchart also informs rigorous reasoning about when narrower, formal algorithmic fairness methods may actually be appropriate.

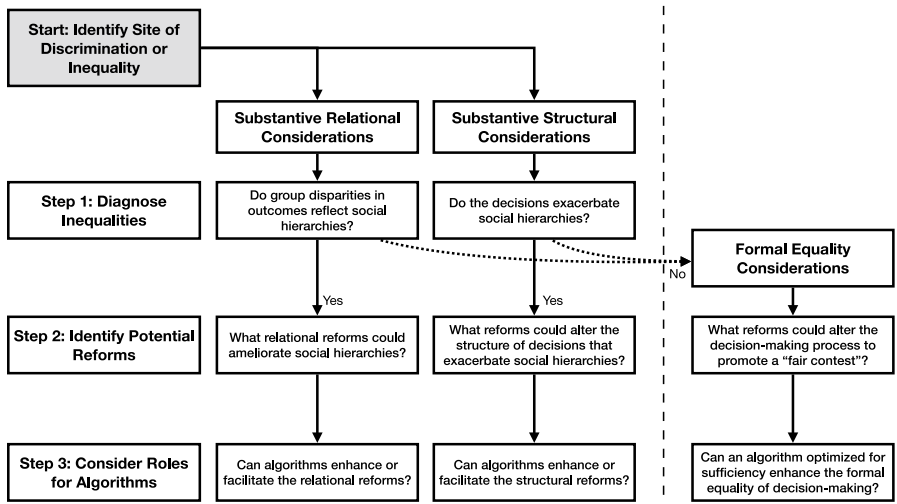

**Fig. 2** Flowchart for implementing substantive algorithmic fairness. The process begins at the top of the flowchart, with concern about discrimination or inequality in a particular decision-making process. This feeds into the substantive equality considerations focused on relational and structural inequalities. If neither relational nor structural concerns are salient (i.e., the answers to both questions in Step 1 are "No"), then the process transitions to formal equality considerations. In this case, the questions resemble those that already exist within formal algorithmic fairness. In this sense, substantive algorithmic fairness represents an expansion of algorithmic fairness methodology rather than a complete rejection of formal algorithmic fairness.

The first step is to diagnose the substance of the inequalities in question. This entails looking for conditions of hierarchy and questioning how social and institutional arrangements reinforce those conditions (MacKinnon, 2011). When faced with disparities in data, substantive algorithmic fairness asks: do these disparities reflect social conditions of hierarchy? Similarly, when faced with particular decision points, substantive algorithmic fairness asks: do these decisions (and the interventions that they facilitate) exacerbate social hierarchies? If the answers to both questions are no, then formal algorithmic fairness presents an appropriate path forward. However, if the answers to these questions are yes—as they often will be when confronting inequalities in high-stakes decisions—then reforms guided by formal equality will be insufficient.

The second step is to consider what types of reforms can remediate the substantive inequalities identified in the first step. Substantive algorithmic fairness draws on the reforms proposed by Anderson (1999), Minow (1991), and Fishkin (2014) for promoting equality without becoming trapped by intractable dilemmas. The first approach is the relational response: reduce dignitary and material disparities that reflect social hierarchies. The relational response attempts to mitigate the upstream social disparities that feed into the decision-making process in question. The second approach is the structural response: alter the structure of decisions to reduce the extent to which decisions exacerbate social hierarchies. The structural response attempts to mitigate the downstream harms that result for those judged unfavorably within the decision-making process in question. Because these reforms target the relational and structural factors that produce equality dilemmas, they provide paths forward that are not subject to the impossibility of fairness.

The third step is to analyze whether and how algorithms can enhance or facilitate the reforms identified in the second step. The critical words here are "enhance" and "facilitate." Rather than treating algorithms as the central component of reform, the analysis here should consider whether and how algorithms can support larger agendas for reform. Thus, in considering the potential role for algorithms, computer scientists should be wary of technological determinism and the assumption that algorithms can remedy all social problems. Algorithmic interventions should be considered through an "agnostic approach" that prioritizes reform, without assuming any necessary or particular role for algorithms (Green & Viljoen, 2020). This approach requires decentering technology when studying injustice and remaining attentive to the broader forces of marginalization (Gangadharan & Niklas, 2019). This analysis will reveal that many algorithms are unnecessary or even detrimental tools for reform. However, this analysis will also reveal new, fruitful roles for algorithms to complement broader efforts to combat oppression.

## 5.2 Applying Substantive Algorithmic Fairness to Pretrial Reform

Substantive algorithmic fairness reveals new strategies for how algorithms can advance pretrial reform. Formal algorithmic fairness suggests that the appropriate pretrial reform strategy is to make release/detain decisions using pretrial risk assessments. In contrast, substantive algorithmic fairness suggests reforms that more robustly challenge the injustices associated with pretrial decision-making and that provide an escape from the impossibility of fairness. Although this approach highlights the limits of pretrial risk assessments, it also suggests new paths for reform and new roles for algorithms.

### 5.2.1 Step 1: Diagnose Inequalities

When pursuing pretrial reform through substantive algorithmic fairness, the first step is to consider the substance of inequalities that manifest in pretrial decision-making. As described in Section 4.1, the disparity in recidivism rates across Black and white defendants reflects a relational harm of racial hierarchy. This disparity cannot be attributed to chance or to inherent group differences (nor is it solely the result of measurement bias). Furthermore, preventative detention presents a structural harm, exacerbating racial hierarchy by depriving high-risk defendants of rights and subjecting them to negative outcomes.

### 5.2.2 Step 2: Identify Potential Reforms

The second step is to consider what reforms could appropriately address the substantive inequalities identified in the first step. Here, we can follow the relational and structural responses. Although these responses are oriented toward reducing incarceration, they do not ignore the group differences in recidivism rates. Instead, the responses inform how to act in light of these disparities, recognizing them as the product of structural oppression. Unlike formal algorithmic fairness, substantive algorithmic fairness neither naturalizes group disparities as neutral facts about the world nor assumes that disparities merely reflect measurement bias. In turn, substantive algorithmic fairness suggests responses to empirical differences in group behaviors that challenge and ameliorate the underlying racial hierarchy.

The relational response suggests altering the relationships that define "risk" to reduce its unequal distribution across the population. This entails interrogating the meaning and distribution of risk as a factor that informs pretrial detention decisions. By recognizing that empirical racial disparities in risk are the product of contingent social arrangements, the relational response provides a strategy for making pretrial adjudication more equitable without ignoring or denying these disparities.

The guiding principle of the relational response is to minimize the link between race and recidivism risk. If risk were not disparately distributed along racial lines (i.e., if Black defendants did not exhibit higher risk than white defendants), then pretrial adjudication would not raise a dilemma between sufficiency and separation. Following this analysis, the relational response suggests reducing the crime risk of Black communities by alleviating criminogenic conditions of disadvantage. For instance, public policies that extend access to education (Lochner & Moretti, 2004), welfare (Tuttle, 2019), and affordable housing (Diamond & McQuade, 2019) all reduce crime, and therefore could reduce the racial disparity in crime risk. In addition, the relational response suggests combatting the association of Blackness with criminality and the effects of this association. This entails not merely challenging stereotypes that link Blackness with crime, but also decriminalizing behaviors that were previously criminalized to subjugate minorities (Butler, 2017; Muhammad, 2011).

The structural response suggests altering the structure of pretrial decision-making to reduce the harmful consequences associated with being high risk to recidivate. This entails responding to risk in ways that go beyond the release/detain binary. Scholars and policymakers assume that more equitable pretrial adjudication stands in direct conflict with public safety (Berk et al., 2018; Corbett-Davies et al., 2017). In contrast, by recognizing that governments can (and should) respond to high-risk defendants without simply detaining them, the structural response provides a strategy for making pretrial adjudication more equitable without compromising public safety.[12]

---

[12] It is worth noting that the relationship between pretrial detention and public safety is not as straightforward as risk assessment proponents typically suggest. Although detention reduces defendants' short-term likelihood of crime, it also increases their long-term propensity for crime, yielding no net effect on future crime (Dobbie et al., 2018). Furthermore, the implementation of pretrial risk assessments has not been shown to reliably reduce crime rates in practice (Stevenson, 2018).

The guiding principle of the structural response is to minimize the link between recidivism risk and punishment. If exhibiting high crime risk did not lead to the loss of freedom and opportunities, then pretrial detention decisions would not exacerbate racial disparities in risk, thus reducing the stakes of the dilemma between sufficiency and separation. Most directly, the structural response can be advanced by reducing the scope of pretrial detention, such that fewer people would be incarcerated, regardless of their risk level. In recent years, for instance, several jurisdictions have stopped pursuing pretrial detention for defendants arrested for misdemeanors and non-violent felonies, without observing any increase in recidivism (Herring, 2020). Another reform would be to begin responding to risk with social and material support, such that being high risk would lead to aid rather than incarceration. Several such programs exist in the USA and have been shown to improve the well-being of defendants while also reducing their recidivism risk (Mayson, 2019). Finally, reforms could aim to decrease the downstream damages of pretrial detention. For instance, reducing the effects of pretrial detention on increased conviction and diminished future employment would reduce the harms associated with being high risk, even if detention remains a common response.[13]

### 5.2.3 Step 3: Consider Roles for Algorithms

The third step is to consider the potential roles for algorithms in advancing relational and structural reforms. Following the relational response, the key question is whether algorithms can enhance or facilitate the identified relational reforms. One direction along these lines involves using algorithms to reduce the crime risk of Black communities by alleviating criminogenic conditions of disadvantage. For instance, algorithms have been used to increase access to education (Lakkaraju et al., 2015), welfare (DataSF, 2018), and affordable housing (Ye et al., 2019), all of which can reduce the crime risk of disadvantaged groups. Another direction involves using algorithms to combat the criminalization of minorities. One such path is to alter the notions of "risk" that guide pretrial decision-making. Pretrial risk assessments define risk in ways that fail to account for the harms of detention on defendants and their communities, which are often more severe for marginalized defendants (Green, 2020; Yang, 2017). Developing more holistic definitions of risk would likely limit the extent to which it appears prudent to detain Black defendants at higher rates than white defendants. More broadly, algorithms can help to undo policies that enact racialized conceptions of crime. Several states have implemented algorithms to streamline the process of expunging criminal records, which is likely to disproportionately benefit minority and low-income individuals (Johnston, 2022). Similarly, statistical analyses have helped to document how stop-and-frisk subjugates minorities and to support movements for altering or abolishing this practice (Denvir, 2015; Goel et al., 2016).

---

[13] As an added benefit, these reforms would reduce the future crime rate of defendants (Dobbie et al., 2018).

Following the structural response, the key question is whether algorithms can enhance or facilitate the identified structural reforms. One path along these lines involves using algorithms to reduce the harms of the racial disparity in recidivism risk. Algorithms can be used to target supportive rather than punitive responses to risk (Barabas et al., 2018; Mayson, 2019), thus mitigating rather than compounding the injustices behind the high recidivism risk of Black defendants. Another direction involves using algorithms to support broader political agendas for structural reforms. For instance, algorithmic evaluations could help justify structural reforms by exposing the false promises of pretrial risk assessments (Angwin et al., 2016; Green & Chen, 2019) and by providing a systemic view of how the criminal justice system exacerbates racial inequalities (Crespo, 2015). Algorithms could also be used to make structural reforms more possible by empowering communities advocating for criminal justice reform and supporting the campaigns of political candidates promising such reforms.

In sum, substantive algorithmic fairness demonstrates how an expansive analysis of social conditions and institutions can lead to rigorous theories of social change, and how those theories of change can inform algorithmic interventions that are not subject to the impossibility of fairness. Starting with these broader reform agendas provides paths for algorithms in pretrial reform that involve more than just pretrial risk assessments. It is important to note that none of these alternative algorithmic interventions would completely solve the problems of pretrial detention—that is an unrealistic goal for any individual reform. Nor are algorithms necessarily the centerpiece of reform. Instead, these algorithmic interventions should be implemented to support broader agendas for substantive pretrial reform. Substantive algorithmic fairness could present similar paths forward in other domains in which the impossibility of fairness has been interpreted as a significant and intractable barrier to reform, such as child welfare (Chouldechova et al., 2018) and college admissions (Friedler et al., 2021).

### 5.3 Substantive Algorithmic Fairness in Practice

Substantive algorithmic fairness offers a new direction for algorithmic fairness. It shifts the field's concern away from formal mathematical models of "fairness" and toward substantive evaluations of whether and how algorithms can combat social hierarchies. Substantive algorithmic fairness therefore expands the tasks involved in promoting equitable public policy with algorithms. Instead of focusing merely on the mathematical properties of algorithms, it is necessary to diagnose the inequalities that are present, evaluate which reforms can best advance substantive equality, and consider how algorithms can support those reforms.

#### 5.3.1 Implications for Practitioners

Substantive algorithmic fairness has significant implications for computer science, the field most centrally focused on algorithmic fairness and the impossibility of fairness. Substantive algorithmic fairness shifts the standards for what it means to make rigorous claims about an algorithm's ability to promote equitable public policy. Under this methodology, it is not sufficient to declare that an algorithm is fair on the basis of mathematical tests alone. Instead, such claims must also account for relational and structural inequities, theories of change for remedying those inequities, and whether algorithms actually represent an effective tool for advancing the desired reforms.

Achieving this methodological evolution will require several shifts in computer science culture and practice. Computer science training must expand beyond its traditional emphasis on the mathematical properties of algorithms to incorporate normative reasoning, sociotechnical systems, and theories of social change. Furthermore, computer science training must inculcate a focus on the real-world impacts of algorithms. In addition to courses focused on ethics and sociotechnical systems, curricula should incorporate practice-based classes in which students collaborate with organizations (e.g., government agencies, nonprofits, and advocacy groups). Such courses can help students consider an algorithm's impacts in light of broader social contexts and appreciate the power of choosing to not design systems that could exacerbate inequality (Graeff, 2020). Advancing substantive algorithmic fairness will also require overcoming cultural and institutional barriers in computer science. The exclusion of women and minorities from algorithm development leads to notions of fairness that are inattentive to the lived realities of oppressed groups (West, 2020). Furthermore, computer science departments disincentivize research that incorporates qualitative modes of reasoning and that prioritizes social impact over novel methodology.

Substantive algorithmic fairness also requires a shift in focus and policy from governments. Many government officials adhere to formal algorithmic fairness reasoning, adopting algorithms on the basis of technical claims about improving decision-making (115th United States Congress, 2017; New Jersey Courts, 2017) and regulating bias as a statistical property of algorithms (116th United States Congress, 2019; Brown, 2020; European Commission, 2021; Government of Canada, 2021; Le, 2021). Because of this limited scope, government uses of algorithms often fail to generate the expected benefits in practice and regulations fail to prevent many algorithmic harms. Policymakers must therefore shift from treating algorithms as technical instruments to recognizing the use of algorithms as a political endeavor. Prior work has argued that choosing a fairness metric and its parameters is a political task that should be made democratically (Wachter et al., 2021; Wong, 2020). Substantive algorithmic fairness demonstrates that democratizing algorithmic fairness requires an even wider scope: it is also necessary to democratize decisions such as how to reform discriminatory policies and whether to use algorithms at all.

Substantive algorithmic fairness therefore centers efforts to improve policy with algorithms around communities, social movements, and reform advocates. These groups typically already follow substantive algorithmic fairness orientations, articulating how seemingly neutral algorithms represent a narrow approach to reform and can entrench unjust policies (Stop LAPD Spying Coalition, 2018; The Leadership Conference on Civil and Human Rights, 2018).

Substantive algorithmic fairness can support these groups with new suggestions for how to incorporate algorithms into their broader visions for structural reform.

### 5.3.2 Implementation Challenges

Even with computer scientists, policymakers, and communities attuned to substantive algorithmic fairness, answering the questions presented by the flowchart in Fig. 2 is a difficult and politically contested task. Substantive algorithmic fairness does not provide a precise roadmap for reform. It presents a sequence of questions, with conceptual tools for answering those questions in a principled manner, rather than a mandatory checklist. This lack of explicit prescription is not so much a limit of substantive algorithmic fairness as an inescapable reality of pursuing substantive social and political reform (Unger, 2005; Wright, 2010). Furthermore, these questions lack neutral and objective answers. Attempts to enact substantive algorithmic fairness will inevitably involve vying with groups that are opposed to structural reforms.

Answering the questions in step 1 will involve grappling with contested notions of what types of inequalities are unjust and what evidence constitutes sufficient proof of social hierarchies. This step represents a significant inflection point within substantive algorithmic fairness, determining whether reform should continue along substantive lines or divert to formal algorithmic fairness. Because there exist both expansive and restrictive views of antidiscrimination (Crenshaw, 1988), simply asking whether social hierarchies are present will not necessarily yield a consensus. Indeed, a central challenge presented by formal algorithmic fairness is that its proponents tend to answer "no" to both questions in step 1, even when analyzing contexts such as pretrial adjudication, in which scholars and communities decry the presence of racial hierarchies. This challenge is exacerbated by the fact that many political actors and technology companies benefit from formal algorithmic fairness, which allows them to embrace "fairness" without making significant political or economic concessions (Bui & Noble, 2020; Green, 2020; Powles & Nissenbaum, 2018). The primary task in step 1 is therefore to ensure that typically disadvantaged communities (along lines of race, gender, class, and so on) have significant voice in determining the answers to the questions. Compared to privileged groups, these communities are generally better able to identify social hierarchies and articulate the contingency of those hierarchies (Collins, 2000; Harding, 1998).

Another challenge in step 1 is to determine the scope of hierarchies under consideration. Throughout this article, I have focused on racial hierarchies. In addition to being widespread and destructive, racial inequity is at the center of debates about the impossibility of fairness. However, social hierarchies exist along many additional lines, including gender, class, ability, sexual orientation, and age (both alongside and in the absence of racial hierarchies). Furthermore, hierarchies manifest in intersecting and interlocking ways (Collins, 2000; Crenshaw, 1989). Thus, when answering step 1's questions about social hierarchies, it is important to consider multiple forms of hierarchy as well as how those hierarchies intersect. One starting point for determining the relevant hierarchies is to focus on the lines along which group disparities in outcomes are most salient (e.g., in the context of pretrial release, disparities in recidivism rates are particularly stark and controversial along racial lines). Analysis should then expand to consider how the most salient hierarchies intersect with additional lines of privilege and disadvantage.

Step 2 requires the difficult task of developing a strategy for remediating the substantive inequities identified in step 1. The relational and structural responses provide general principles for identifying potential reforms. However, even with these principles in hand, there is no single or straightforward path for how to achieve change. One of the central challenges in advancing social change is determining which reforms to pursue in any specific situation, among many potential paths forward (Unger, 2005; Wright, 2010). Sociologist Erik Olin Wright provides one strategy for reducing a broad set of reforms into a more concrete and actionable set (Wright, 2010). First, focus on desirability: develop a wide list of alternative policies and social arrangements that are normatively appealing, at least in the abstract. Second, focus on viability: evaluate which of the desirable alternatives would likely mitigate social hierarchies if implemented in practice, based on available knowledge about institutional designs and social organizations. Third, focus on achievability: consider which of the viable reforms could actually gain the support and traction necessary to be implemented.

Many reforms that are desirable and viable will not be achievable, at least within the short term. This could suggest that substantive algorithmic fairness is merely a practice of utopian idealism. However, substantive algorithmic fairness does not assume or require that all of its suggested reforms are immediately achievable. Instead, it presents a slate of options, which reformers can choose from based on the avenues that appear most promising within the particular terrain of contestation. Confronting the challenges of achievability reveals that what may appear to be a weakness of substantive algorithmic fairness—its lack of a precise roadmap for reform—is actually a strength.

It is impossible to identify a concrete reform strategy that is achievable in every circumstance. Achievability is highly contingent on the particular actors, policies, and ideologies at play within a given domain (Wright, 2010). Substantive algorithmic fairness therefore eschews specific blueprints in favor of principles that suggest a multitude of complementary and modular reforms. The relational and structural responses complement each other but are beneficial even if enacted in isolation. Furthermore, these responses each suggest many potential reforms, all of which can be beneficial in isolation. Thus, even though substantive algorithmic fairness begins with an ambitious vision of substantive equality, it is oriented toward guiding incremental reforms for achieving this goal. What substantive algorithmic fairness provides, in other words, is a method for rigorously developing incremental reforms that push public policy in the direction of greater substantive equality. In this sense, substantive algorithmic fairness takes after political theories of "real utopias" (Wright, 2010), "non-reformist reforms" (Gorz, 1967), and prison abolition (Davis, 2003), all of which present strategies for distilling long-term, radical agendas for social justice into short-term, piecemeal reforms.

The first consideration in step 3 is whether algorithms can advance the reforms identified in step 2. Computer science pedagogy, corporate advertising, and the media promote the idea that technology provides solutions to social issues (Broussard, 2018; Green, 2019; Morozov, 2014). However, Step 3 requires accepting that algorithms may not be productive tools for promoting certain reforms at all, and at minimum cannot achieve most reforms on their own. Algorithms can best help to promote social change when deployed in conjunction with broader policy and governance reforms (Abebe et al., 2020; Green, 2019). One strategy for deciding whether and how to use algorithms without falling into solutionism is to engage with communities advocating for reform. Recent work provides several examples of how data analysis and technology design can be incorporated into community-driven reform efforts that challenge oppression (Asad, 2019; Costanza-Chock, 2020; Lewis et al., 2018; Maharawal & McElroy, 2018; Meng & DiSalvo, 2018).

The second consideration in step 3 is whether an algorithm that appears beneficial will actually promote the desired impacts in practice. Algorithmic interventions are indeterminate, often leading to impacts that differ from what was expected based on technical evaluations of the model (Green & Viljoen, 2020). For instance, pretrial risk assessments have not generated the expected increases in pretrial release, in large part because judges respond to the algorithmic recommendations in punitive and racially biased ways (Albright, 2019; Stevenson, 2018). One strategy for anticipating the downstream impacts of a given algorithm is to run pre-deployment experiments that test how people interact with its advice (Green & Chen, 2021). Such experiments can provide proactive knowledge about whether the algorithm is likely to improve decision-making in practice and whether alternative design mechanisms can improve human–algorithm collaborations.

## 6 Conclusion

Algorithmic fairness provides an increasingly influential toolkit for promoting equitable public policy. It is therefore essential to consider whether algorithmic fairness provides suitable conceptual and practical tools to guide reform. Because formal algorithmic fairness restricts analysis to isolated decision points, it cannot account for social hierarchies and the impacts of decisions informed by algorithms. As a result, formal algorithmic fairness traps reform efforts within the impossibility of fairness and suggests reforms that uphold social hierarchies. Substantive algorithmic fairness provides a new orientation for algorithmic fairness, incorporating algorithms into broader agendas for reform. In doing so, substantive algorithmic fairness offers an escape from the impossibility of fairness and suggests new roles for algorithms in combatting oppression.

Nonetheless, even as substantive algorithmic fairness provides tools for escaping the impossibility of fairness, it does not provide an escape from all normative conflict. To start with, there is no single definition of the normative principles animating substantive algorithmic fairness. Concepts such as social justice are "essentially contested," and as such resist any singular, undisputed definition (Gallie, 1955). Egalitarian thinkers have long debated what exactly should be made equal across people (e.g., luck, money, capabilities, relationships) and to what population equality applies (e.g., all humans, all residents of a particular jurisdiction, all legal citizens of a particular jurisdiction) (Arneson, 2013). Substantive algorithmic fairness cannot settle these debates, but it is aligned most closely with relational egalitarian theories.

Moreover, even for those aligned around broad understandings of substantive equality, advancing their vision requires jockeying with people and institutions committed to maintaining social hierarchies. Substantive algorithmic fairness describes one slice of terrain on which fights for a more egalitarian society should be waged; it cannot ensure that egalitarian forces will have sufficient power to achieve all of their goals. Thus, as with all efforts to achieve substantive equality, substantive algorithmic fairness requires ongoing political struggle to achieve conditions amenable to reform.

Finally, the principles of justice and equality that underlie substantive algorithmic fairness do not stand on their own as sole or supreme values. Both within and across societies, many essential values coexist: justice, equality, liberty, loyalty, compassion, virtue, dignity, and so on. Although these values often align, they also unavoidably clash and cannot be resolved into a clear hierarchy (Berlin, 2013). Substantive algorithmic fairness does not comprehensively capture all normative ends, and its pursuit must be balanced with other societal considerations. What substantive algorithmic fairness provides instead is a method for determining how algorithms can promote equitable public policy, particularly in light of decision-making processes that raise concerns about discrimination and injustice.[14] These are the goals that motivate much of the technical research and regulation on algorithmic fairness (Berk et al., 2018; Booker, 2019; Kleinberg et al., 2019; Le, 2021), even though the existing tools of formal algorithmic fairness are ill-equipped for the job.

Although substantive algorithmic fairness does not yield a precise or comprehensive roadmap for reform, it provides a compass to help computer scientists and policymakers reason about the appropriate roles for algorithms in combatting inequity. Debates about algorithms often feature a binary contest between algorithmic reforms and the status quo: when critics challenge the use of algorithms, proponents argue that the only alternative to implementing fallible and biased algorithms is to rely on fallible and biased humans (Berk et al., 2018; Kleinberg et al., 2019; Miller, 2018). Substantive algorithmic fairness demonstrates that reformers need not accept this binary choice between implementing a superficially "fair" algorithm and leaving the status quo in place. Instead, there are many potential reforms to consider—all of them, in some form, incremental—and many potential roles for algorithms to enable or supplement those reforms. By starting from substantive accounts of social hierarchy and social change, the field of algorithmic fairness can stitch together incremental algorithmic reforms that collectively build a more egalitarian society.

---

[14] The relational emphasis of substantive algorithmic fairness also aligns algorithmic fairness more closely with values such as compassion and dignity.

**Acknowledgments** I am grateful to Elettra Bietti, Matt Bui, Ben Fish, Evan Green, Will Holub-Moorman, Lily Hu, Abbie Jacobs, Andrew Schrock, Salomé Viljoen, Zach Wehrwein, and the reviewers for valuable suggestions on how to improve this manuscript. I thank the Michigan Society of Fellows for feedback and financial support.

**Author Contribution** BG conceived of and wrote the entire manuscript.

## Declarations

**Conflict of Interest** The author declares no competing interests.